\documentclass[%
 reprint,
 amsmath,amssymb,
 aps,
]{revtex4-2}
\usepackage{xcolor}
\usepackage{float}
\usepackage{graphicx}
\usepackage{dcolumn}
\usepackage{bm}

\usepackage{pdfpages}
\makeatletter
\AtBeginDocument{\let\LS@rot\@undefined}
\makeatother

\begin{document}
\preprint{APS/123-QED}

\title{Towards predictive atomistic simulations of SiC crystal growth}

\author{Alexander Reichmann$^a$}
\email{Corresponding author: alexander.reichmann@unileoben.ac.at}
\affiliation{$^a$Department of Materials Science, Christian Doppler Laboratory of Advanced Computational Design of Crystal Growth, Montanuniversität Leoben, Leoben, Austria\\$^b$Ebner European Mono Crystalline Operations (EEMCO) GmbH, Leonding, Austria\\$^c$ present address: Infineon Technologies Austria AG, Siemensstrasse 2, Villach, Austria}

\author{Zahra Rajabzadeh$^a$}%
\affiliation{$^a$Department of Materials Science, Christian Doppler Laboratory of Advanced Computational Design of Crystal Growth, Montanuniversität Leoben, Leoben, Austria\\$^b$Ebner European Mono Crystalline Operations (EEMCO) GmbH, Leonding, Austria\\$^c$ present address: Infineon Technologies Austria AG, Siemensstrasse 2, Villach, Austria}

\author{Sebastian Hofer$^b$}
\affiliation{$^a$Department of Materials Science, Christian Doppler Laboratory of Advanced Computational Design of Crystal Growth, Montanuniversität Leoben, Leoben, Austria\\$^b$Ebner European Mono Crystalline Operations (EEMCO) GmbH, Leonding, Austria\\$^c$ present address: Infineon Technologies Austria AG, Siemensstrasse 2, Villach, Austria}

\author{René Hammer$^{b,c}$}
\affiliation{$^a$Department of Materials Science, Christian Doppler Laboratory of Advanced Computational Design of Crystal Growth, Montanuniversität Leoben, Leoben, Austria\\$^b$Ebner European Mono Crystalline Operations (EEMCO) GmbH, Leonding, Austria\\$^c$ present address: Infineon Technologies Austria AG, Siemensstrasse 2, Villach, Austria}

\author{Lorenz Romaner$^a$}%
\affiliation{$^a$Department of Materials Science, Christian Doppler Laboratory of Advanced Computational Design of Crystal Growth, Montanuniversität Leoben, Leoben, Austria\\$^b$Ebner European Mono Crystalline Operations (EEMCO) GmbH, Leonding, Austria\\$^c$ present address: Infineon Technologies Austria AG, Siemensstrasse 2, Villach, Austria}

\date{\today}

\begin{abstract}

Simulations of SiC crystal growth using molecular dynamics (MD) have become popular in recent years. They, however, simulate very fast deposition rates, to reduce computational costs. Therefore, they are more akin to surface sputtering, leading to abnormal growth effects, including thick amorphous layers and large defect densities. A recently developed method, called the minimum energy atomic deposition (MEAD), tries to overcome this problem by depositing the atoms directly at the minimum energy positions, increasing the time scale.

We apply the MEAD method to simulate SiC crystal growth on stepped C-terminated 4H substrates with 4° and 8° off-cut angle. We explore relevant calculations settings, such as amount of equilibration steps between depositions and influence of simulation cell sizes and bench mark different interatomic potentials. The carefully calibrated methodology is able to replicate the stable step-flow growth, which was so far not possible using conventional MD simulations. Furthermore, the simulated crystals are evaluated in terms of their dislocations, surface roughness and atom mobility. Our methodology paves the way for future high fidelity investigations of surface phenomena in crystal growth.

\end{abstract}

\maketitle

\section{\label{sec:level1}Introduction}

The wide band gap material silicon carbide (SiC) has gained more and more importance for the realization of electronic devices, due to its unique properties, which enable operation at elevated frequencies, voltages, and temperatures. These properties allow SiC to experience a wide range of use-cases, which range from power devices \cite{SHUR1998}, the automotive industry \cite{Wrzecionko2014}, to quantum computing \cite{son2020} and quantum sensing \cite{castelletto2022}.  While SiC has several advantages over Si and is believed to replace it in many applications, its complex and time consuming crystal growth process leads to much higher production costs. SiC crystals are usually grown at very high temperatures, between 2200-2800 K, and mainly via the physical vapor transport (PVT) method, which has much smaller growth rates compared to growth from the melt. It is difficult to track the thermodynamic processes that occur in the furnace during growth \cite{Fornari2018}, and therefore our understanding of the detailed conditions of growth and the formation of defects is still limited \cite{LIU2019}. One type of defect, the stacking fault in particular, is of great relevance, since SiC can occur in more than 200 different polytypes \cite{JEPPS1983,XU2021}, with different electrical and mechanical properties. Especially the band gap of the most commonly occurring polytypes 3C, 6H and 4H is 2.4, 3 and 3.2 eV, respectively \cite{Wellmann2018}, showing considerable variation. Understanding the conditions that favor the growth of a certain polytype is crucial, and computational studies can provide deep insights.\\

Polytype stability has been investigated to a large extent using ab-initio methods \cite{Karch1994,Park1994,mercier2012,ITO2013,Limpijumnong2017} and compared to experimental results  \cite{Inomata1968,Heine1991,YAKIMOVA2000}. 

The experimental investigations show that the polytype stability depends heavily on the temperature. At lower temperatures 3C is generally formed \cite{Heine1991}, while 4H and 6H are favored at elevated temperatures \cite{YAKIMOVA2000}. In addition, the polytype grown is also influenced by the surface's characteristics \cite{KIMOTO2016}, the addition of doping elements \cite{SCHMITT2008} or other factors, such as the C/Si ratio in the growing surroundings \cite{Vodakov1982}. 

DFT investigations at 0K, also don't paint a clear picture, since the free energy difference between the polytypes is in the range of a few meV/atom, which is small compared to the energy difference induced by the choice of exchange-correlation functional. For example, using the generalized gradient approximation (GGA), 4H and 6H are lower in energy than 3C at 0K. Adding van der Waals corrections can lead to 3C being the most stable polytype, but not for every implementation \cite{Kawanishi2016}. Dou et al. \cite{DOU2011} have reported that, looking at the phonon interactions, 4H becomes the most stable polytype, while 3C is metastable. Scalise et al. \cite{Scalise2019} have analyzed the stability at elevated temperatures via calculations of vibrational entropy and found that 4H and 6H are more stable than 3C, in agreement with experiment. Similar to the experiment, it has also shown in DFT calculations that doping changes the polytype stability \cite{NISHIZAWA2019,rajabzadeh2025}, especially for donors since the conduction band of the polytypes is energetically strongly shifted.  

Of great importance are surface phenomena that may allow a polytype to grow, even if its bulk free energy is unfavorable. Recently Ramakers et al. \cite{Ramakers2022} have calculated surface energetics and concluded that, while 4H and 6H are favored in the bulk phase, 3C has a lower surface energy which may favor it during growth at lower temperatures. Direct simulations of crystal growth phenoman on the surface are, however, computationally too demanding for DFT. In the past, such simulations have mostly been carried out using kinetic Monte Carlo (kMC) methods \cite{Cheimarios2021}, including super lattice kMC \cite{CAMARDA2007,camarda2013,Fisicaro2020} or competitive lattice kMC \cite{GUO2015,huang2016}. While these methods capture important processes such as step-bunching \cite{LI2016,CHEN2019} or island growth \cite{AI2025}, they rely on predefined deposition sites, that remain fixed during the growth simulation. This makes it difficult to realistically simulate dislocation behaviour during growth. Furthermore, the important processes taking place on the surface and their reaction rates must be input to the calculation and the corresponding choices may drastically determine the outcome of the simulation.\\
 
In recent years atomistic investigations of SiC crystal growth have increasingly been carried out via molecular dynamics (MD) simulations \cite{KANG2014,XUE2022,kangli2023,Kayang2025}, using interatomic potentials (IPs). While no constraints on crystal structure or defects are imposed in such simulations, they have substantial computational costs compared to kMC, result in smaller length scales (a few nm) and shorter timescales (a few ns). Therefore, deposition rates in the simulations exceed the experimental values by far ($\sim10^7$-$10^9$ times higher). This may be the reason why previous MD simulations \cite{KANG2014,kangli2023} have revealed thick amorphous layers and predominant growth of 3C on top of a [0001] oriented 4H substrate, even at high temperatures. So far MD methods were not able to replicate the kinetic stabilization of a 4H polytype using a stepped substrate, as it is commonly used in experiments \cite{MATSUNAMI1997,KIMOTO2016}. Therefore, the real growth mechanism is not captured by such simulations.\\    

Promising approaches to combine the advantages of MD and MC have been proposed. Most of these are hybrid approaches \cite{Neyts2013}, such as time-stamped force biased MC (tfMC), which combines the force calculation of MD with the hopping probability of MC. Another method was recently proposed by Karewar et al. \cite{Karewar2024} and is called the minimum energy atomic deposition (MEAD) method. It works by first scanning over the potential energy landscape (PEL) of a system and finding minimum energy positions (MEPs).  Afterwards, atoms are deposited in these MEPs. Before the next scanning of the PEL is commenced, the system can be equilibrated by tfMC or, as it was proposed in the original work, using the fast inertial relaxation engine (FIRE) \cite{GUENOLE2020} . The high versatility of the method makes it ideally suited to explore growth phenomena for many systems. We will show in this work that MEAD can realistically capture the growth mechanism of 4H SiC, overcoming the shortcomings of both MD and MC. We describe how the MEAD algorithm can be adapted to SiC and reveal step-flow phenomena as well as the occurrence of crystal defects.

\section{Methodology}
\label{sec:Theory}

The general form of the MEAD method, as described by \cite{Karewar2024} implements the following steps: 
\begin{enumerate}
    \item Calculate the PEL of the surface via a grid search of potential deposition sites (also called phantom atoms by the original work)
    \item Remove potential deposition sites which don’t fulfill the minimum energy condition 
    \begin{equation}
    \label{eqn:energy_cutoff}
    |\frac{E_i-E_{min}}{E_{min}}|<E_{cut-off},
    \end{equation}
    where $E_{cut-off}=0.15$ was chosen by \cite{Karewar2024}
    \item Remove low energy deposition sites which are in too close proximity to respective lower energy deposition sites
    \item Populate all remaining deposition sites 
    \item Apply additional minimization steps, such as the FIRE algorithm or tfMC at pre-defined temperature values
    \item Repeat the steps above until a sufficiently large amount of atoms are deposited
\end{enumerate}

With this method Karewar et al. \cite{Karewar2024} were able to demonstrate crystal growth simulations of Si deposited on Si, Al deposited on Si and Al deposited on Al, which are in good agreement with experiment. For the deposition of one atom species it seem to work reasonable well, however for the co-deposition of two species, changes to the algorithm are necessary, in addition to other modification specific for the SiC System.

\begin{figure}
  \centering
  \includegraphics[scale=0.5]{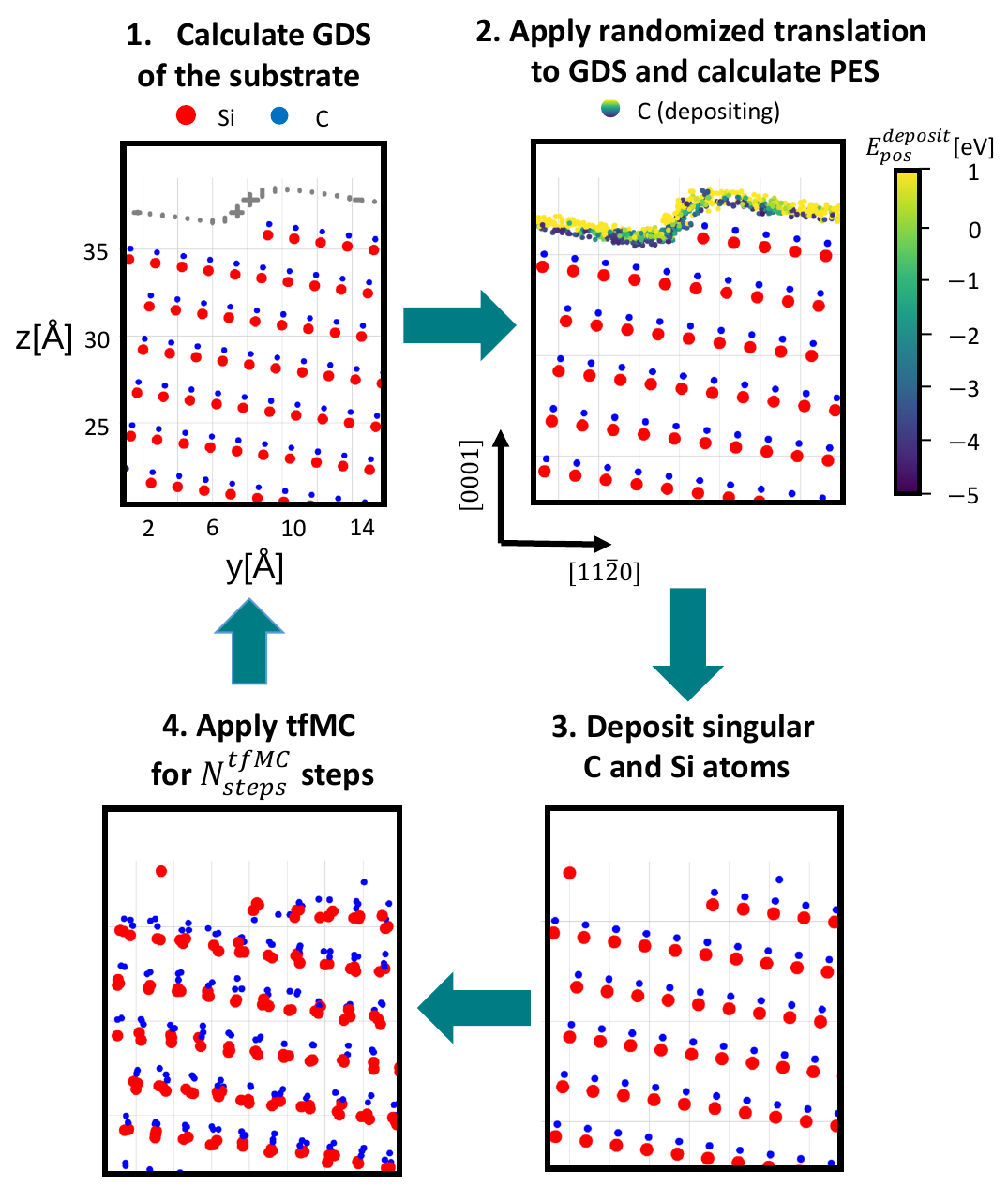}
  \caption{The MEAD algorithm as we applied it onto the SiC system.}
  \label{fig:MEAD_steps}
\end{figure}

Our implementation of the MEAD method is the following. A schematic of the MEAD algorithm, step by step, as it was implemented for the SiC system, is given in Fig.\ref{fig:MEAD_steps}. 

\begin{enumerate}
    \item Calculate the PEL via the Gaussian density surface (GDS) method \cite{Krone2012} : We use the GDS, which was already adopted in  \cite{Karewar2024} but for different purpose, to generate potential deposition sites. This is illustrated in Fig.\ref{fig:MEAD_steps} on a stepped surface, where the mesh of the GDS perfectly follows the stepped morphology of the surface, approximately 1 \AA \, above the surface. Similarly to \cite{Karewar2024} each potential deposition site from the mesh is then randomly displaced to gain a better estimate of the PEL. For the PEL the most important influence comes from the IP. PELs calculated by the most commonly used IPs of SiC have been compared to DFT calculations and are presented in the supplemental informations. For all further calculations we have chosen the MEAM IP from Kang et al. \cite{KANG2014}, especially because it has been used for MD crystal growth simulations multiple times before \cite{KANG2014,kangli2023,Kayang2025}. 
    
    \item Populate the PEL: Since the original MEAD algorithm was designed for the deposition of one atom type we modified the algorithm to accommodate the co-deposition of both Si and C. The minimum energy condition for removal (step 2 of the original MEAD algorithm) can cause issues if the energy range of deposition sites of one atom type is lower than that of the other, which may lead to non-stoichiometric deposition, or in the worst case, suppression of the deposition of one atom type. Therefore, the potential energy surface is evaluated first for the C atom and one C atom is deposited. Then the PEL is evaluated for Si for the surface with the added C and one Si atom is deposited. Since we target the lowest possible growth rate we do not take into account the option to deposit more than one atom. 
    
    To select the deposition site the following procedure is adopted. All potential energy sites that do not fulfill the energy cut-off criteria (Eq. \ref{eqn:energy_cutoff}) are removed. Then a site is chosen randomly, to which a probability weighting is applied according to the distance in [0001] direction by using the form $ w=(1-r_d)^{\Delta z-1} \text{ for } \Delta z >1 \text{\AA}$, where $\Delta z$ is the distance from the lowest position found by the GDS and $r_d$ is the decay rate. This parameter can be tuned to either  choose sites close to the surface more aggressively (if $r_d$ is close to 1), or allow more broad deposition (if $r_d$ is close to 0). This prevents artifacts such as the deposition on atoms in the gas phase (after desorption) or the formation of nano-pillars (further details are given in the supplemental informations).

    \item Equilibration of structure: Since the grid search of the PEL in the last step only gives an estimate for the location of the MEP an additional equilibration step is necessary. In the original work by Karewar et al. \cite{Karewar2024} the FIRE energy minimization \cite{GUENOLE2020} was used. In addition, the possibility of using tfMC \cite{Neyts2013} was explored. Since this is closer to reality for high temperature applications we adapted the second approach. For our simulation we set the maximum displacement distance of the tfMC approach $\Delta=0.2$ \AA, which is slightly higher than the 0.15 \AA \, of Karewar et al. \cite{Karewar2024}. Multiple $\Delta$ values ranging from 0.1 to 0.3 \AA \ were tested, and 0.2 \AA \, has been chosen as the best compromise. 
    \item Repeat the steps above until a sufficiently large amount of atoms are deposited
\end{enumerate}

To implement the MEAD algorithm two softwares have been used. For the calculation of the PEL and the tfMC LAMMPS \cite{PLIMPTON1995} have been used, while for the GDS and the selection of deposition sites a Python script was used, which originates from \cite{Karewar2024}, but which was modified for our purpose.
For the identification of the crystal structure and the dislocation analysis, the open-source visualization tool OVITO \cite{Stukowski2010} was used. The crystal structure was determined with the "identify diamond structure" modification, which uses the method of \cite{MARAS2016}. The dislocation analysis uses the dislocation extraction algorithm from \cite{Stukowski2010_dis}. Both these modification are well established in the field of material science \cite{Gruber2017,Zhang2020,kangli2023,Kayang2025} 

\section{Results and Discussion}
\label{sec:thermo}

\begin{figure*}
  \centering
  \includegraphics[scale=0.23]{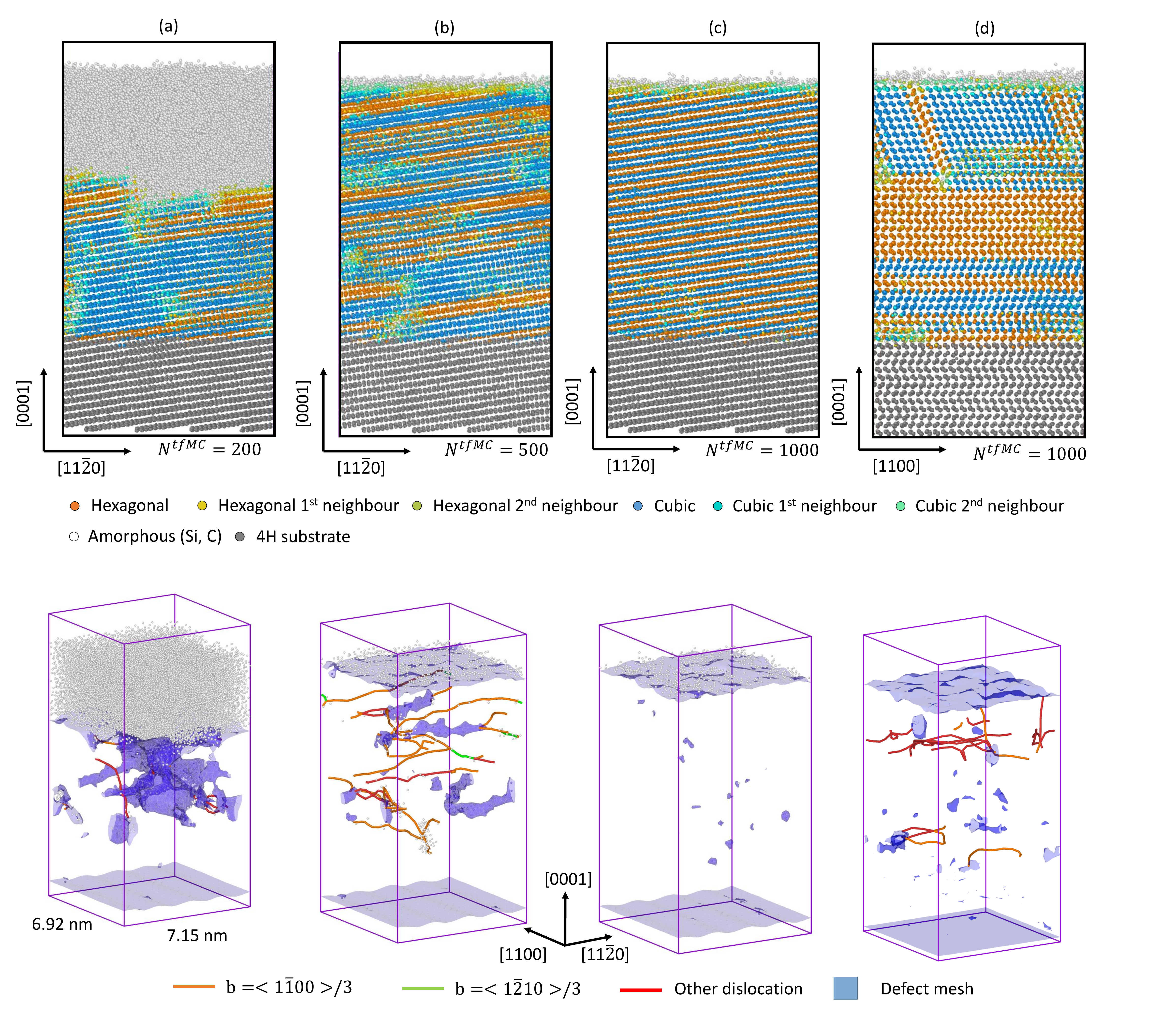}
  \caption{The MEAD algorithm applied on the 7$\times$7 stepped (a-c) and flat (d) 4H C-terminated SiC substrate. The amount of tfMC steps in between Si and C depositions was varied between (a) 200, (b) 500 and (c,d) 1000 tfMC steps. The dislocation analysis of the respective final structure is depicted below. The "identify diamond structure" modification of OVITO classifies each atom into one of the following categories with the associated color in brackets: hexagonal diamond (orange), hexagonal diamond 1$^{\text{st}}$ neighbor (bright yellow),  hexagonal diamond 2$^{\text{nd}}$ neighbor (lime green), cubic diamond (blue), cubic diamond 1$^{\text{st}}$ neighbor (aquamarine), cubic diamond 2$^{\text{nd}}$ neighbor (bright green) and "other" (white). Here 1$^{\text{st}}$ neighbor indicates that four neighboring atoms are positioned on the lattice, but at least one of its second nearest neighbors is not and 2$^{\text{nd}}$ neighbor indicate that the atom itself is on the lattice, but at least one of its neighbors is either missing or not positioned on a lattice site. The substrate was set to a grey color.}
  \label{fig:crystal_growth_small}
\end{figure*}

For the crystal growth simulations we started with a 7.1 nm $\times$ 6.9 nm 4H substrate, which will be called 7$\times$7 substrate in the following. The stepped variant of the substrate are described first, since we calibrate the amount of tfMC steps needed and other MEAD parameters by the ability to stabilize the 4H substrate through step-flow growth. With this cell size a perfectly repeating stepped substrate with an off-cut angle of 8.1° was made. The height of this stepped substrate is 3.1 nm. The substrate contains a total amount of 14664 atoms and for each crystal growth simulation 40000 atoms, 20000 Si and 20000 C atoms, were deposited. The tfMC temperature was set to 2500 K and the decay rate of the probability weights to $r_d=$0.5. 

\begin{figure*}
  \centering
  \includegraphics[scale=0.33]{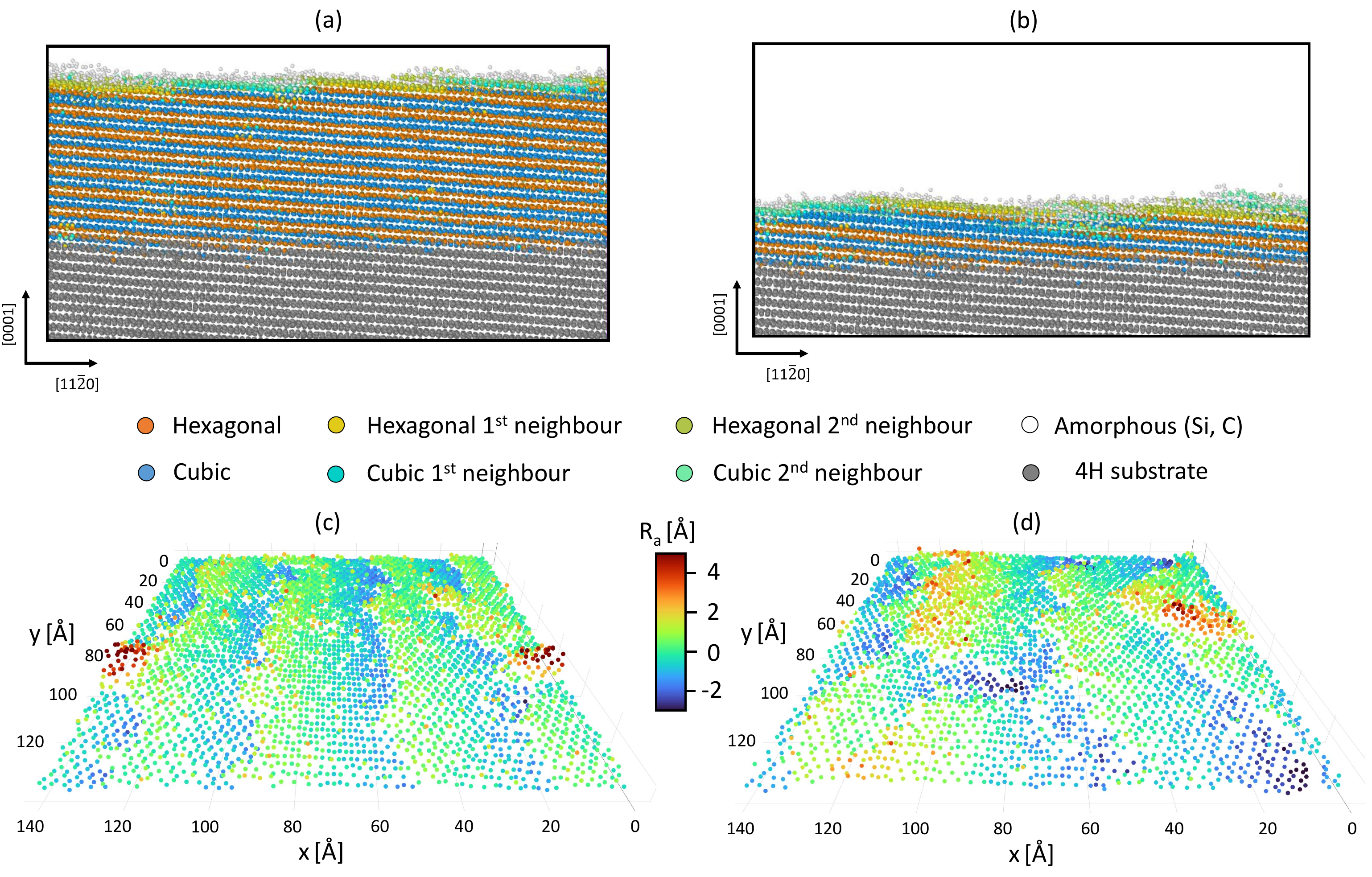}
  \caption{The MEAD algorithm applied on a  14$\times$14 stepped substrate using (a) 4000 and (b) 8000  tfMC steps in between depositions. For (a) a larger decay rate of the probability weights was applied, meaning the deposition was forced to be closer to the steps. In (a) a total of 80000 atoms have been deposited while in (b) only 26000 atoms were deposited before the simulations was stopped. The evaluation of their respective surface roughness R$_a$ are depicted in (c) and (d), in a three-dimensional perspective, with the viewpoints inclined towards the [000$\bar{1}$] direction. For the structure (a) a dislocation density analysis is shown in (e) and an analysis of the movement of 14 randomly picked C and Si atoms over the first 15000 MEAD steps is shown in (f). Videos of the step flow growth of both simulations are provided as supplemental materials (in which 1 frame is roughly 200 MEAD steps). They can be viewed under DOI:10.5446/72185 for the simulation using 4000 tfMC steps and DOI:10.5446/72186 for the simulation using 8000 tfMC steps.}
  \label{fig:crystal_growth_big}
\end{figure*}

Our calibration concluded that, at least 1000 tfMC steps (Fig.\ref{fig:crystal_growth_small}c) are necessary between each deposition for step-flow growth to stabilize the 4H structure of the substrate, while with 500 steps (Fig.\ref{fig:crystal_growth_small}b) large amounts of stacking faults are still present in the substrate and at lower deposition rates (200 steps, Fig.\ref{fig:crystal_growth_small}a) a large amorphous layer is still present. The fact that crystal defects appear at higher growth rates is consistent with experiment \cite{OHNO2004}. We are not aware of any experimental evidence that has reported amorphous layers during growth. It may be due to deposition rates that are not achievable in experiment. 

The structure which exhibits step-flow growth also appears to be dislocation free, according to the dislocation analysis of OVITO. Only point defects, such as interstitial atoms can be seen. Using also 1000 tfMC steps and the same MEAD parameter for a crystal growth simulation of a flat C-terminated 4H substrate (Fig.\ref{fig:crystal_growth_small}d), shows a large amount of stacking faults. There, clusters of atoms in either cubic or hexagonal formation, so only 3C and 2H SiC are forming, while there is no indication for 4H, 6H or other larger polytypes. The ratio of cubic to hexagonal is almost half, which is in contrast to conventional MD simulations, which in past mostly have seen 3C \cite{KANG2014,kangli2023} in the past. However predominant 2H growth using the MEAM IP have also recently been reported \cite{Kayang2025}. Indeed, deposition of 2H is puzzling, since 2H has the lowest stability of all the main polytypes, predicted by the MEAM IP, at least for the bulk phase. Ramakers et al. \cite{Ramakers2022}, however already showed for ab-initio that surface energies can be quite different and even predicted low surface energies for 2H.  

It is not straightforward to assign a unique deposition rate to the calculations just presented. Indeed, also the original work of Karewar et al. \cite{Karewar2024} has not provided detailed informations in this respect. The deposition rate is controlled in step 2 and step 3 of the MEAD algorithm. The time of step 3 is well defined since the statistical time associated to 1 tfMC step was calculated to be 6.35 fs by using 

\begin{equation}
\langle t_{tfMC} \rangle=\frac{\Delta}{3}\sqrt{\frac{\pi m_{min}}{2 k_BT}},
\end{equation}

where $m_{min}$ is the mass of the smallest element in the simulation (C) \cite{Neyts2013}. For the lowest deposition rate, we have therefore 6.35 ps per deposition of an Si/C atom pair and a total deposition length of 127 ns. The height of the final structure is about 12 nm, therefore in 127 ns around 9 nm were deposited corresponding to 250 m/h. This is still much higher than experimental deposition rates of 100 $\mu$m/h. However, also step 2 of the MEAD algorithm should be considered which controls the deposition of new atoms. On the one hand, the deposition positions are preselected in this step which mimics the time that it takes an atom, that lands randomly somewhere on the surface, to find the energetically favorable positions via diffusion. Due to the diffusion analysis below, the time scale of this process is not order of magnitudes higher, but probably in the range of about 1000 tfMC steps. Still by applying this procedure the computational effort is reduced by approximately 50 \%. On the other hand, and even more importantly, in the PVT process the supply of gas atoms is controlled via the flux of the species through the chamber from source to seed. In order to replicate these conditions, the unrealistic high number of 10$^9$ tfMC steps would have to be carried out to wait until the next atom is added in step 2. It is reasonable to assume that the surface will remain in a stable unchanged equilibrium state for most of the time span between the addition of the new atoms. Therefore, once enough time is given to the system to equilibrate, and we see this is the case at 1000 tfMC steps, the outcome of the simulation can be expected to be very similar for slower deposition rates. While our simulation may not provide a perfectly equilibrated state due to omission of rare adsorption-desorption events, the uncertainties related to the interatomic potential discussed in the previous section are probably much more relevant to be considered for improving the fidelity of the simulations. We leave such improvements to further investigations in future.

Next we investigated substrates of the size 14.2 nm $\times$ 13.8 nm, which will be called 14$\times$14 substrates, for simplicity. Since they are double in size compared to the 7$\times$7 substrate the angle of the steps is 4.05°. The height of the substrate was increased to 5 nm. Using the same setup as the previous MEAD simulations, we were not able to stabilize the 4H structure through step-flow growth (Fig.\ref{fig:crystal_growth_big}b) even with a very large amount of tfMC steps of 8000 in between depositions. However, when the probability of deposition near the surface was enhanced by setting the decay rate of the probability weights to $r_d= $0.9, it was possible to stabilize the 4H structure through step-flow growth with 4000 tfMC steps  (Fig.\ref{fig:crystal_growth_big}a) , which calculates to a growth rate of around 20 m/h. 


The surface roughness evidences the shape of the surface steps and was evaluated using $R^a_i=z_i-\langle z_{surface} \rangle$, where $\langle z_{surface} \rangle$ is the average z position of all surface atoms. As shown in Fig.\ref{fig:crystal_growth_big}c, steps are straight and evenly spaced for the simulation were 4H grows regularly. In contrast, the step front is curved for the structure exhibiting stacking faults (Fig.\ref{fig:crystal_growth_big}d). This is in good agreement with KMC studies from literature \cite{Krzyżewski2014,CHEN2025}, which report growth of straight steps for higher temperature and lower growth rate. In addition, slight step bunching of two steps can also be observed  (Fig.\ref{fig:crystal_growth_big}d upper left corner), which is complemented by simulations provided in the supplemental informations also showing a correlation between step bunching and stacking faults.

Experimentally, a link between step bunching and stacking faults has previously been discussed by Kollmuss at el. \cite{Kollmuss2023}. Similarly, Wang et al. \cite{WANG2025} found step bunching and surface irregularities such as groves leading to higher stacking fault density and Ohtomo et al. \cite{ohtomo2017} saw wavy step morphology when nucleation of a stacking fault occurs at the growth front, for nitrogen doped PVT growth.  

In our growth simulation slight nucleation growth can also be observed (Fig.\ref{fig:crystal_growth_big}c), even though the decay rate of the probability weights for this run was set at 0.9 (meaning the deposition is more forced towards the steps). Dislocation analysis of the structure revealed, similarly to the 8.1° off-cut structure that, when 4H is stabilized through step-flow growth no dislocation can be identified and only point defects are seen (see supplemental informations).

\begin{figure}
  \centering
  \includegraphics[scale=0.32]{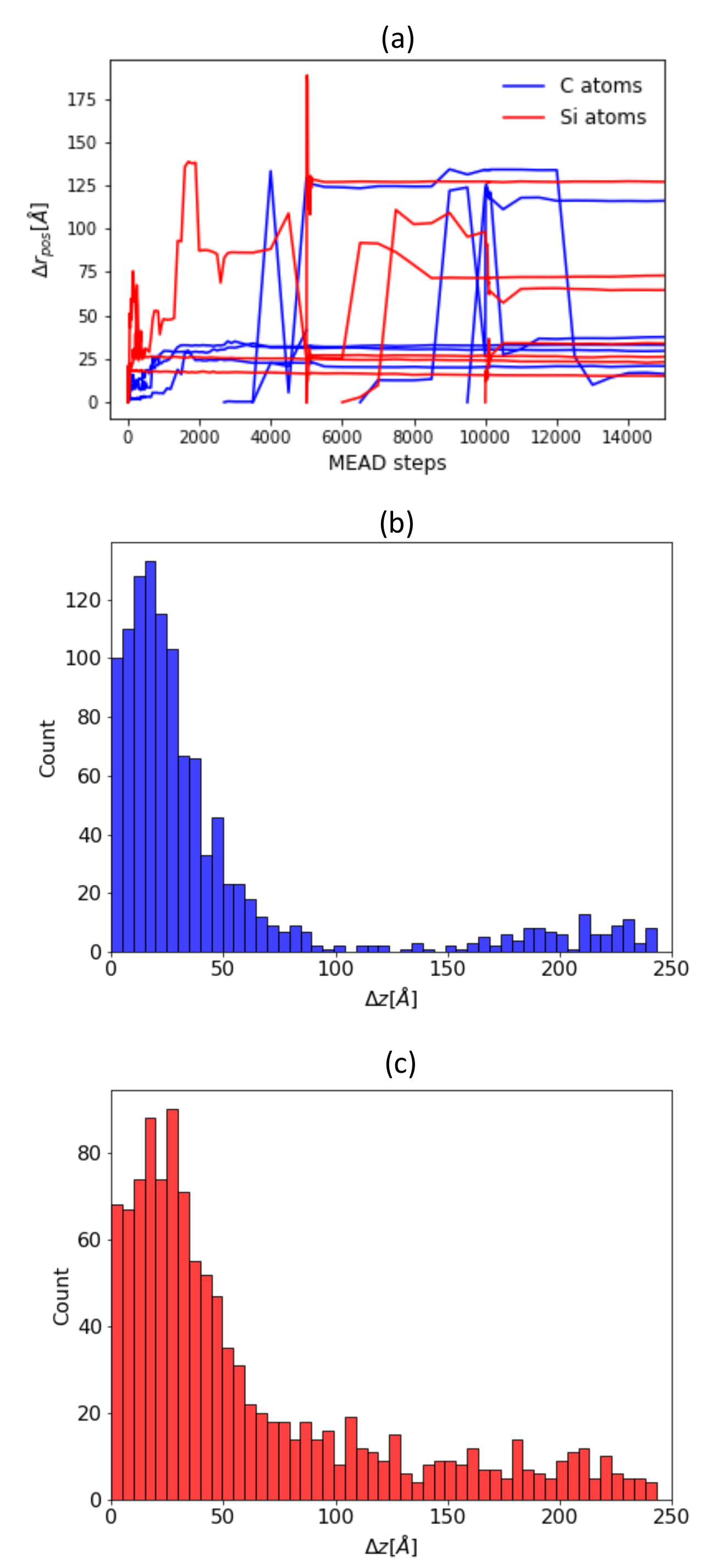}
  \caption{(a) An analysis of the movement of 14 randomly picked C and Si atoms over the first 15000 MEAD steps. (b) and (c) A statistic of the first 1000 deposited C (b) and Si (c) atoms in terms of their difference in final position compared to their deposition position.}
  \label{fig:atom_movement}
\end{figure}

In addition, 7 random Si and C atoms were tracked (Fig.\ref{fig:atom_movement}a) in terms of their movement in relation to the deposition position ($\Delta r_{pos}=\sqrt{\Delta x^2+\Delta y^2+\Delta z^2}$), sampled over the first 15000 MEAD steps from the crystal growth simulation using 4000 tfMC steps in between depositions (Fig.\ref{fig:crystal_growth_big}a). None of the atoms investigated stayed close to its deposition position and the atoms move between 2 nm to 17.5 nm away from their respective initial deposition. The time until the atoms find a stable position varies quite a lot between the ones analyzed. With some finding a stable position within 500 MEAD steps while one particular Si atom took more than 10000 MEAD steps to find its stable position. Although the aforementioned Si atom seems to have found a stable position for several periods of time, which would indicate diffusion behavior within the deposited bulk crystal. A difference between the Si and C atoms in terms of their movement can not clearly identified (Fig.\ref{fig:atom_movement}b and c). A statistic of the difference between deposition position to final position of the first 1000 deposited Si and C atoms show that while most atoms stay roughly 2-5nm away from their respective deposition site, there is also a slight bump of atoms (especially for C atoms), which move up to 20 nm away from their deposition site.

\subsection{Comparison to MD}
\label{sec:thermo}

\begin{figure}
  \centering
  \includegraphics[scale=0.2]{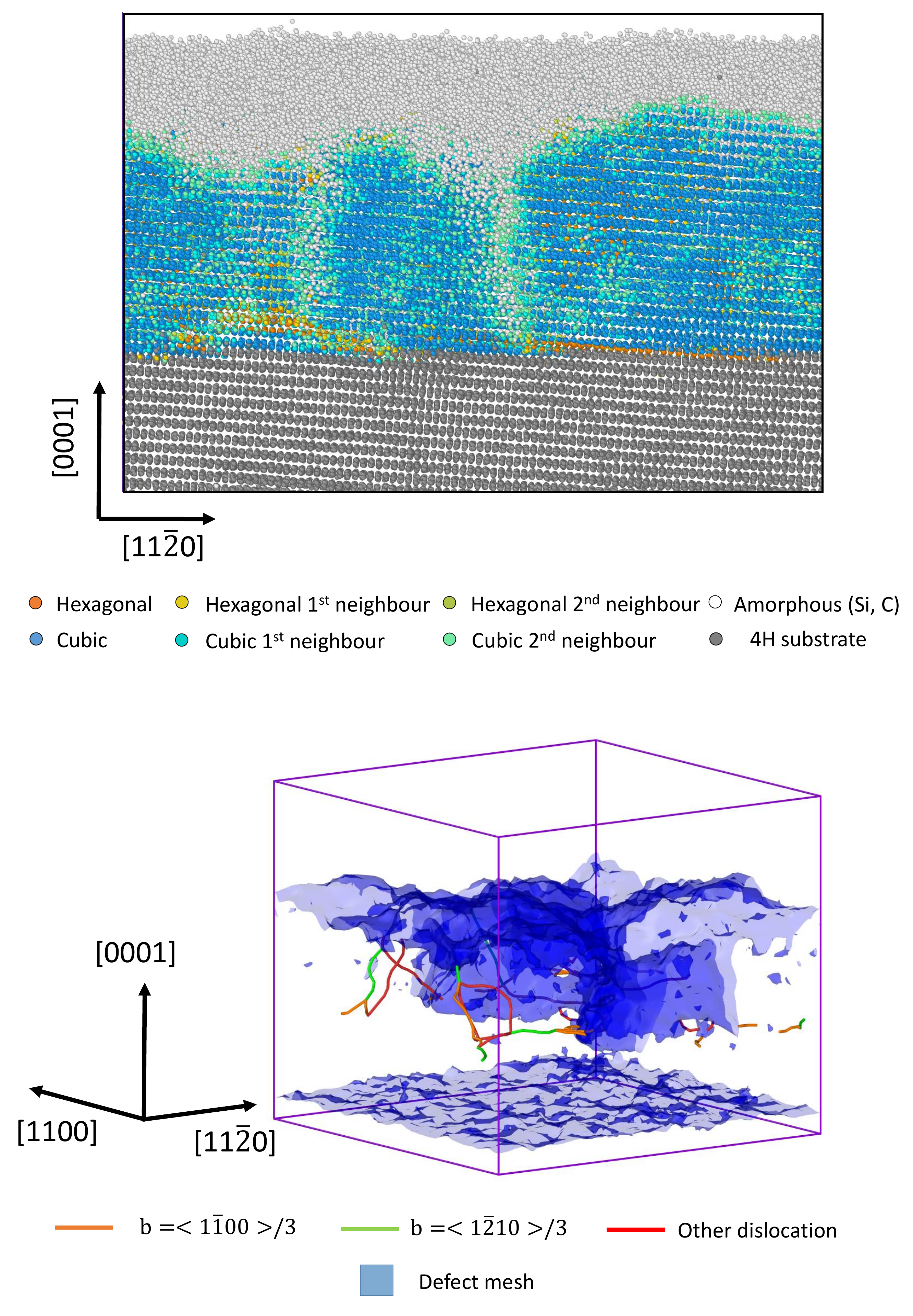}
  \caption{Conventional MD crystal growth simulation applied on the 14$\times$14 stepped 4H substrate with. The dislocation analysis is depicted below.}
  \label{fig:comparison_MD}
\end{figure}

As comparison, the same stepped substrates have been used to conduct conventional MD, using a setup similar to simulations from literature \cite{kangli2023,Kayang2025}. For these, every 1 ps Si and C atoms were inserted in the vacuum above the substrate at a distance of 200 \AA, with an initial velocity in z direction specified by $v_d=\sqrt{\frac{k_BT}{m}}$, where m is the mass of the inserted atom. The temperature of the whole system was kept constant using a Nosé-Hoover thermostat (NVT) \cite{nose1984,Nose1984_2,Hoover1985,Shinoda2004}. The total simulation time was 15 ns for the 7$\times$7 substrate and 40 ns for the 14$\times$14 substrate. The final structure, after deposition on the 14x14 substrate is depicted in Fig.\ref{fig:comparison_MD}, while the one of the 7$\times$7 substrate, which gives similar results, is in the supporting information. A spring constant of 10 eV/\AA \, has been applied to the bottom two layers of the substrates in order to keep their position during the growth simulation. This is needed because the depositing atoms would move the entire substrate downwards over time. To check if this constraint does not introduce additional effects, the same constraint was also imposed for the MEAD simulations with the 7$\times$7 substrate. It was observed that in this case a slightly higher number of tfMC steps were necessary to stabilize the step-flow growth.

\begin{figure*}
  \centering
  \includegraphics[scale=0.48]{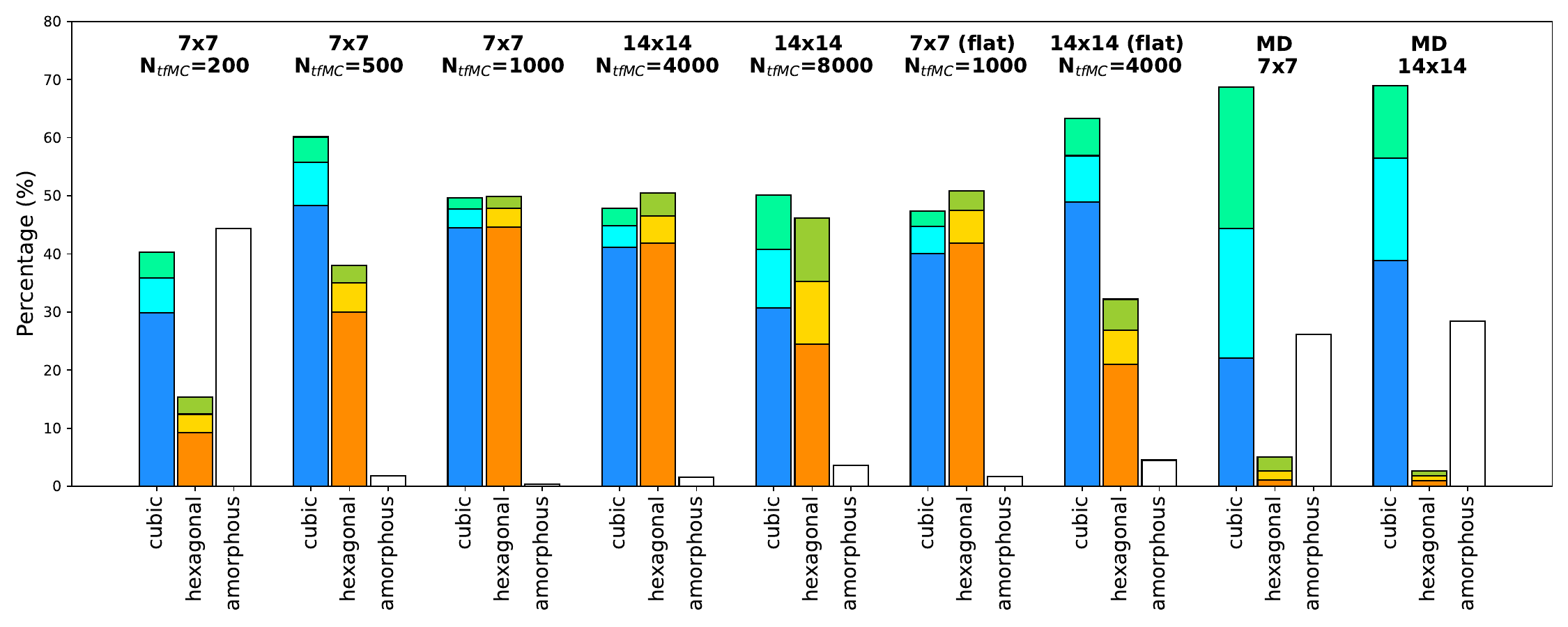}
  \caption{The percentage of crystal structures of the deposited layers of the MEAD and MD simulation of the various substrates and parameters. The same color scheme was used as explained in Fig.\ref{fig:crystal_growth_small}. }
  \label{fig:percentage_crystal}
\end{figure*}

The final structure of the conventional MD crystal growth simulation shows a predominant cubic (3C) crystal structure, in the same vein as previously reported by Kang et al. \cite{KANG2014} and Kangli et al. \cite{kangli2023}. The stepped surface was not able to stabilize the 4H structure, for both the 8.1° and 4.05° off-cut substrates. Especially the 14$\times$14 substrate experiences a large amorphous layer which leads to an elevated number of defects within the crystal. Looking at the crystalline structure of all the simulations conducted (Fig.\ref{fig:percentage_crystal}) shows this as well. This is also in agreement with the MEAD simulation of the 7$\times$7 substrate using only 200 tfMC steps (fast deposition rate), however there are still relatively moderate amounts of hexagonally configured atoms present, in addition to a larger amorphous layer. The crystal structure of the MEAD simulation on the flat substrate also favors cubic on the larger 14$\times$14 structure, while it slightly favors hexagonal on the smaller 7$\times$7 structure.

\section{Conclusion}
\label{sec:conclusion}

The minimum energy atomic deposition algorithm has successfully been applied to the SiC system and clear advantages over conventional MD and MC have been demonstrated. With it, step-flow growth on both 8.1° and 4.05° off-cut stepped C-terminated 4H substrates has been simulated and the 4H structure of the substrate could be stabilized for the entire growth, at a temperature of 2500 K. On flat substrates a mixture of both cubic 3C and hexagonal 2H has formed. The shape of the steps during step-flow growth was tracked and uneven step growth and step bunching seem to be linked with the formation of stacking faults, which is in good agreement to experimental results from literature. In addition the movement of Si and C atoms have been investigated and it could be seen that both types of atom have a tendency to move relatively far away from their original deposition sites. Diffusion within the deposited crystal could also be identified, which shows the strength of this approach. However, the large amount of tfMC steps necessary indicates the challenges of the SiC specific conditions, which limits the system sizes, which can be looked at. Nevertheless, the ability to simulate step-flow growth shows the potential the MEAD algorithm has and especially paired with a modern machine learned interatomic potential it will open the door for important investigations, such as the growth processes changes considering doping elements or other inclusions or crystal growth simulation on screw dislocations.

\section{Acknowledgments}
\label{sec:acknowledgments}

This work was funded by the Christian Doppler Research Association and the Ebner European Mono Crystalline Operations (EEMCO) GmbH. The computational results have been achieved using the HPC facility of the Technical University of Leoben. In addition, we want to thank Germain Clavier (IUT Grand Ouest Normandie, France) for sharing his code, regarding the MEAD algorithm and for his support.  

\section{Declaration of competing interest}
\label{sec:comp_int}

The authors declare that they have no known competing financial interests or personal relationships that could have appeared to influence the work reported in this paper.

\bibliographystyle{unsrtnat}
\bibliography{sample}

\clearpage
\section*{Supplemental informations}


\section*{Supplementary material}

# SUPPLEMENTAL INFORMATIONS

## S1. Crystal structure of SiC polytypes

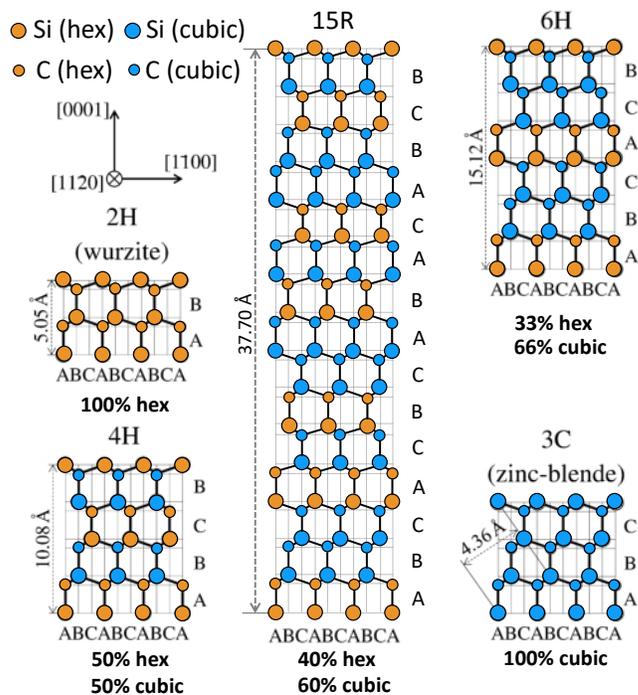

Figure S1: The crystal structure of the 5 most common polytypes.

## S2. Substrates used for crystal growth simulations

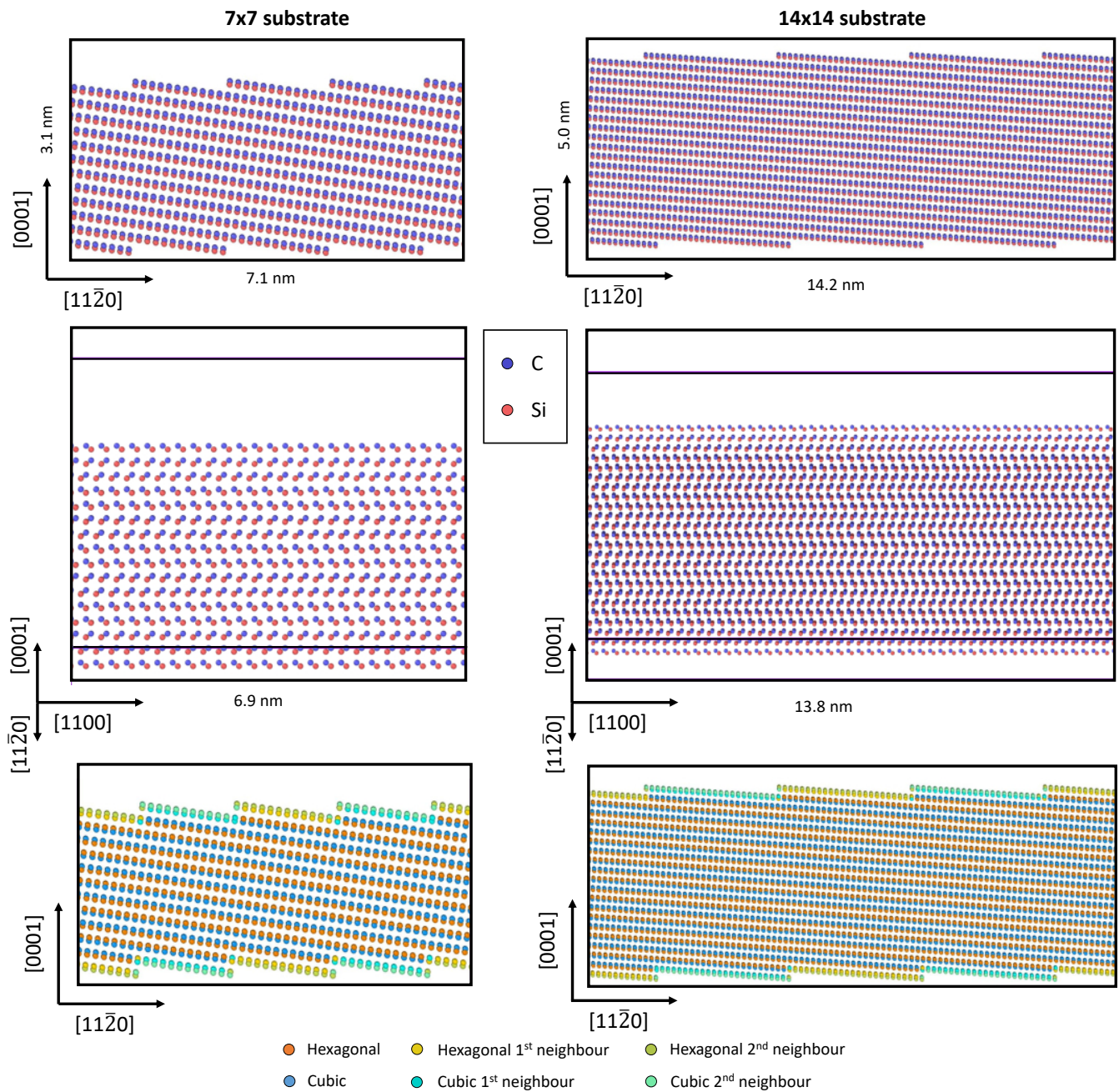

Figure S2: Depiction of the stepped 7x7 (left) and 14x14 (right) substrates used for the crystal growth simulations. Below is the crystal structure of the respective substrate.

2

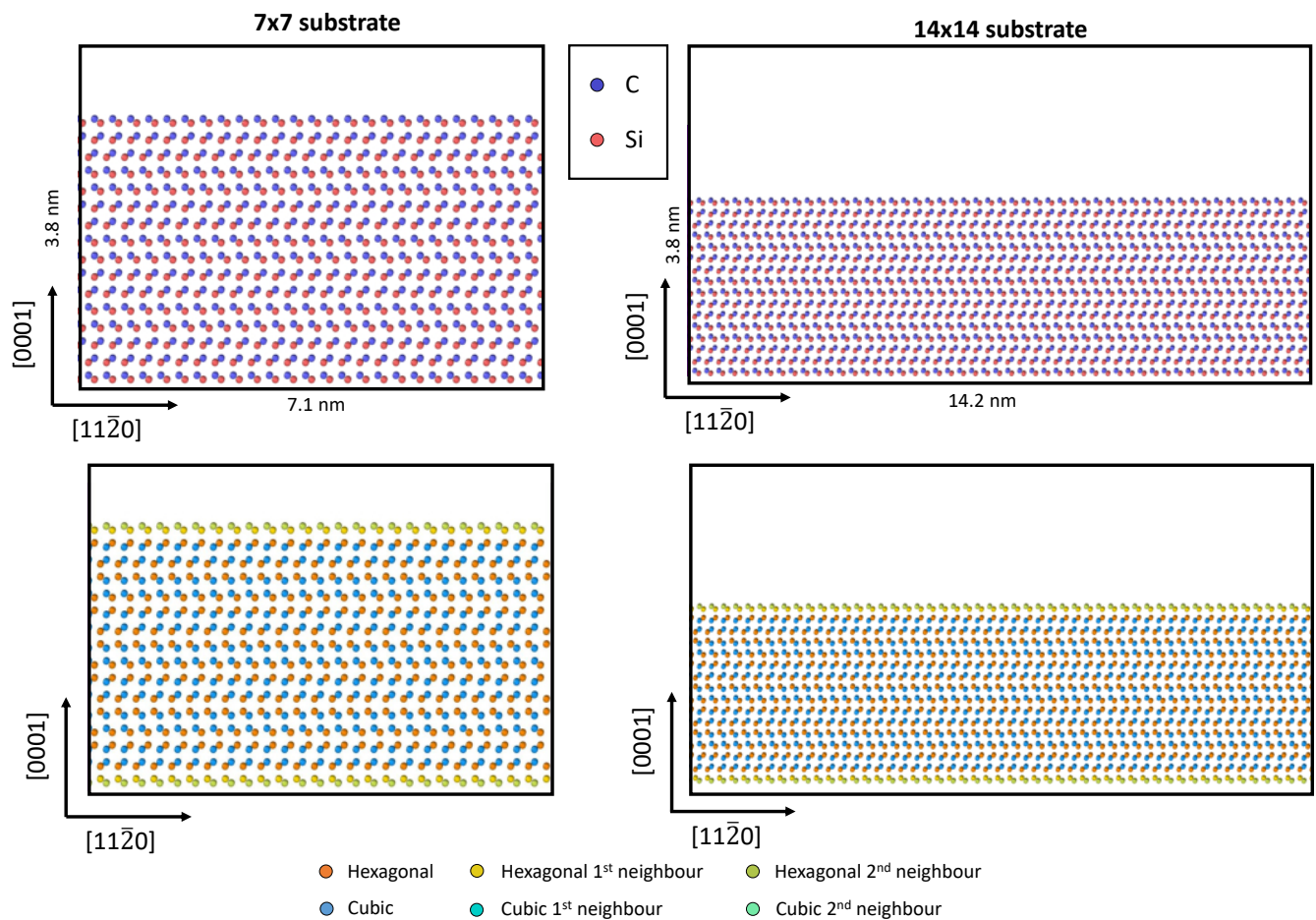

Figure S3: Depiction of the flat 7x7 (left) and 14x14 (right) substrates used for the crystal growth simulations. Below is the crystal structure of the respective substrate.

## S3. Evaluation of the potential energy landscape with interatomic potentials

For the potential energy landscape (PEL) the most important influence comes from the interatomic potential (IP). It is therefore unavoidable to evaluate the available IPs for SiC. There have been done a multitude of comparisons in the past, for many different factors [1, 2, 3, 4], however not for the PEL. We have made an estimate of the PEL using the five most widely used semi-empirical IPs, the MEAM from Kang et al. [5], the environment dependent IP (EDIP) from Jiang et al. [6], the IP from Vashishta et al. [7] and two Tersoff potentials, the T94 [8] and the T05 [9]. In addition, one newer machine learning IP (MLIP), the GRACE universal potential [10] has also been used. Here the FS-OAM GRACE potential has been used, after testing various GRACE potentials on the SiC system. As a comparison the PEL has also been evaluated using DFT.

For this we calculated the energy change when placing singular C and Si atoms on top of a (12Å× 11Å× 18Å) 4H [0001] SiC substrate on different deposition sites (on a grid), without geometric optimization. The first principle calculation were preformed utilizing the Vienna ab initio simulation package (VASP) [11, 12, 13], employing the GGA exchange correlation potential from Perdew et al. [14] (PBE). For the uniform k-mesh, the Monkhort-Pack scheme [15] was applied with a grid size of 6x6x1 and the plane wave cut-off energy was set to 400 eV. In Fig.S4, the PEL of both C (left) and Si (right) atoms on a C-terimanted 4H substrate is compared for DFT, MEAM and GRACE. The PEL of the T05, EDIP, T94 and Vashishta is shown in Fig.S5. The MEAM IP, seems to have a relatively unusual shape, and the PEL drops to 0 comparatively close to the surface of the structure. The GRACE MLIP, visually, agree better with the DFT results, however it slightly underestimates the energy values on top of the substrate, while overestimating the energy levels in the interlayer planes. Looking at the PEL of the DFT, MEAM IP and GRACE MLIP, the fact that in the interlayer plane low energy positions can be found is of concern for the MEAD algorithm. Because to find the MEPs Karewar et al. [16] conducted a grid search in close proximity to the surface. This may lead to interstitial depositions in the case of SiC, especially for uneven surfaces. For this reason the Gaussian density surface (GDS) method is used for the generation of potential deposition sites. The other four IP, namely, the T05, EDIP, T94 and Vashishta are not able to predict low energy positions in the interlayer plane, and the energy values of their depositions sites are generally quite low compared to the DFT. Another concern arises looking at almost all calculated PELs (except the MEAM). They all show a quite uniformal distribution of energy values along the [1100] axis, with no clear indication of valleys with MEPs. This would indicate that the mobility of Si and C atoms atop the surface is high and in order for a crystal structure to be formed a large amount of neighboring atoms and a large amount of time steps may be necessary.

For a better comparison of all the IPs in terms of their agreement with DFT data we used two metrics, the adsorption height $h_{min}$ and the energy value $E_{min}$ of the minimum energy position and comprised them in Tab.S1.

Table S1: A comparison of the most commonly used IPs of SiC with DFT results for adsorption height $h_{min}$ and energy value $E_{min}$ of the MEP of a C and Si atom ontop of a C-terminated 4H substrate.

| type | C adsorption | | Si adsorption | |
|---|---|---|---|---|
| | $E_{min}^{surface}$ | $h_{min}$ | $E_{min}^{surface}$ | $h_{min}$ |
| DFT | −6.5 eV | 1.2 Å | −5.9 eV | 1.7 Å |
| MEAM | −9.3 eV (−5.4 eV*) | 0.2 (0.9*) Å | −2.4 eV | 0.9 Å |
| EDIP | −1.2 eV | 1.5 Å | 0.7 eV | 1.5 Å |
| T94 | −0.3 eV | 1.2 Å | −0.2 eV | 1.9 Å |
| T05 | −0.8 eV | 1.4 Å | −0.5 eV | 1.9 Å |
| Vashishta | −1.5 eV | 2.2 Å | −2.4 eV | 2.4 Å |
| GRACE | −801 eV (−6.7 eV*) | 0 (0.7*) Å | −4.3 eV | 0.4 Å |
| * $E_{min}$ and $h_{min}$ at least 0.5 Å away from surface | | | | |

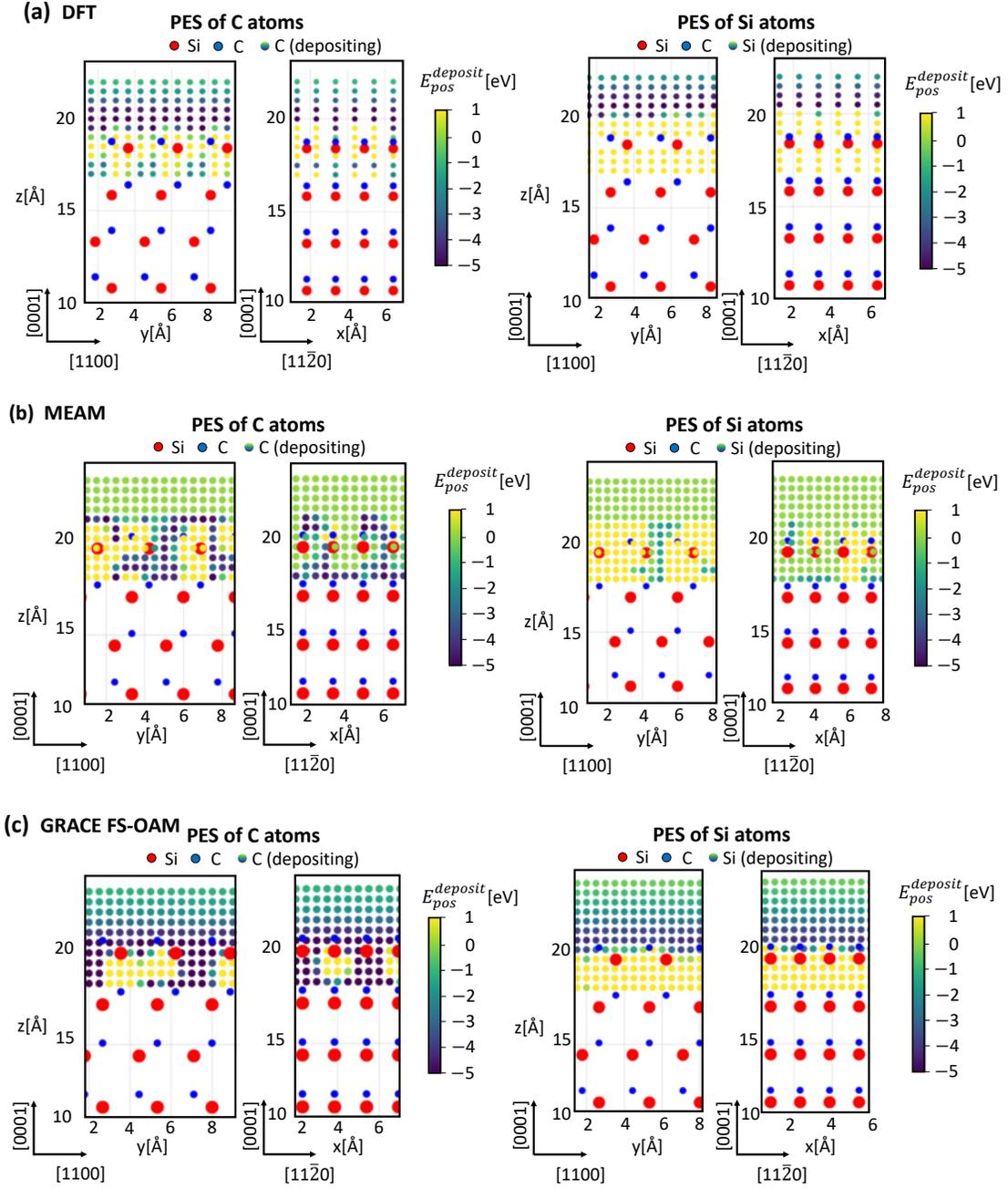

Figure S4: The potential energy landscape of (a) DFT, (b) the MEAM potential and (c) the GRACE FS-OAM potential.

The adsorption height of the MEAM IP and the GRACE MLIP are rather low compared to DFT. In fact the GRACE potential appears to have troubles, when two C atoms are too close together, where it predicts binding instead of repulsion (hence the very low $E_{min}^{surface} = -801$ eV). When only looking at deposition sites at least 0.5 Å away from the substrate (as it would be generated by the GDS) they agree better with the DFT results. The adsorption heights of the T94, the Vashishta and the EDIP potential are in better agreement but the energy values of the MEP are significantly smaller, as could already be seen in Fig.S5. From all the IPs looked at the Vashishta is the only one that does not allow C-C and Si-Si bonding, which can be seen by its diatomic energy curves from NIST database on IPs [17].

To conclude this segment, the GRACE MLIP best agrees with the DFT results. The distribution of MEP above the surface is similarly uniformal with values in close range to DFT. The binding energy also slowly decays going away from the surface and it is able to predict interlayer MEPs, even though it overestimates their energy values. However, in order to have a better comparability to previous studies of MD crystal growth simulations of 4H SiC [5, 18, 19], which were conducted using the MEAM IP, we opted to use the MEAM IP for all further crystal growth simulations.

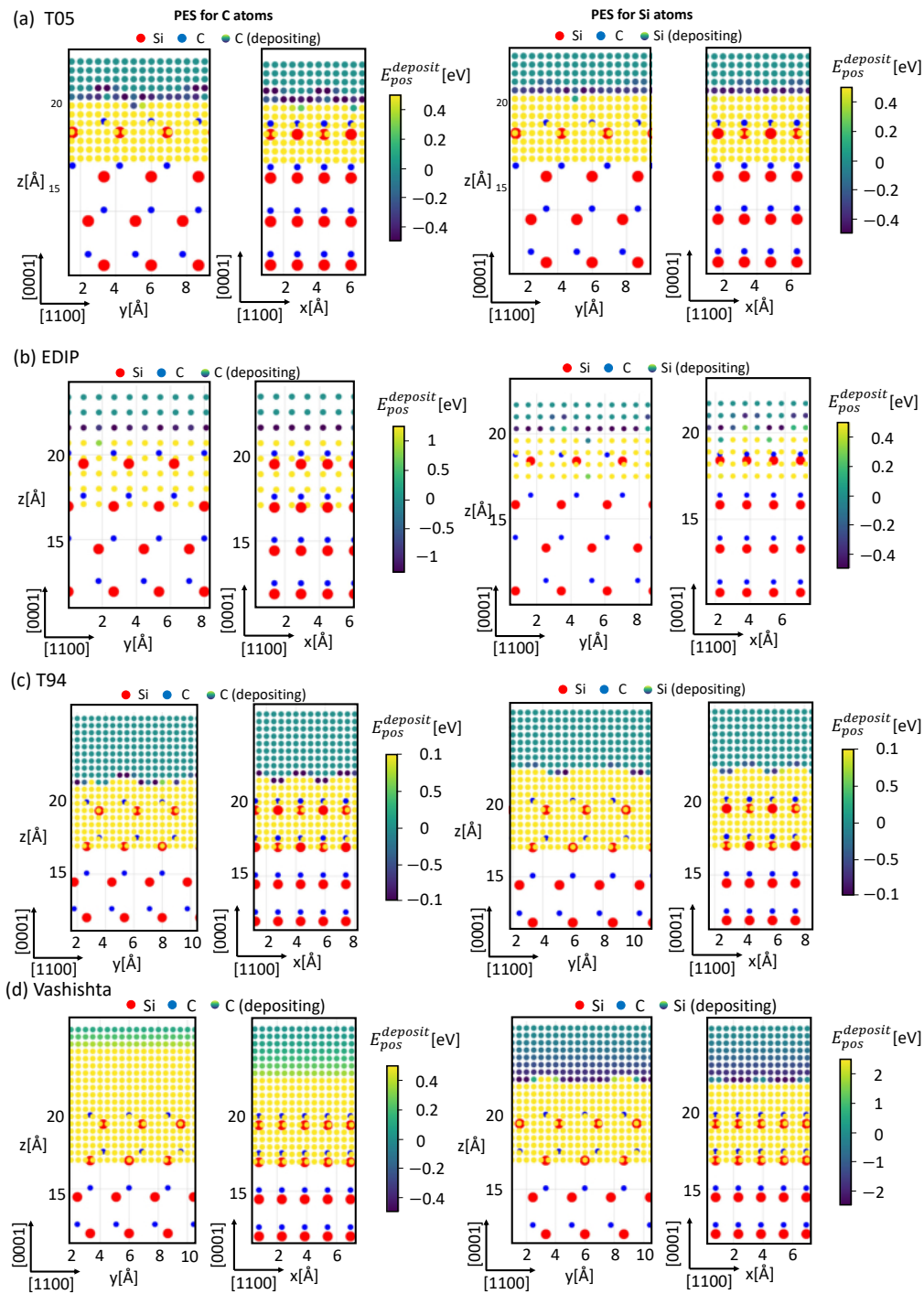

Figure S5: A comparison of the PEL of the (a) T05 (b) EDIP (c) T94 and (d) Vashishta IP on a flat 4H C-terminated SiC substrate.

7

## S4. Minimum energy atomic deposition algorithm

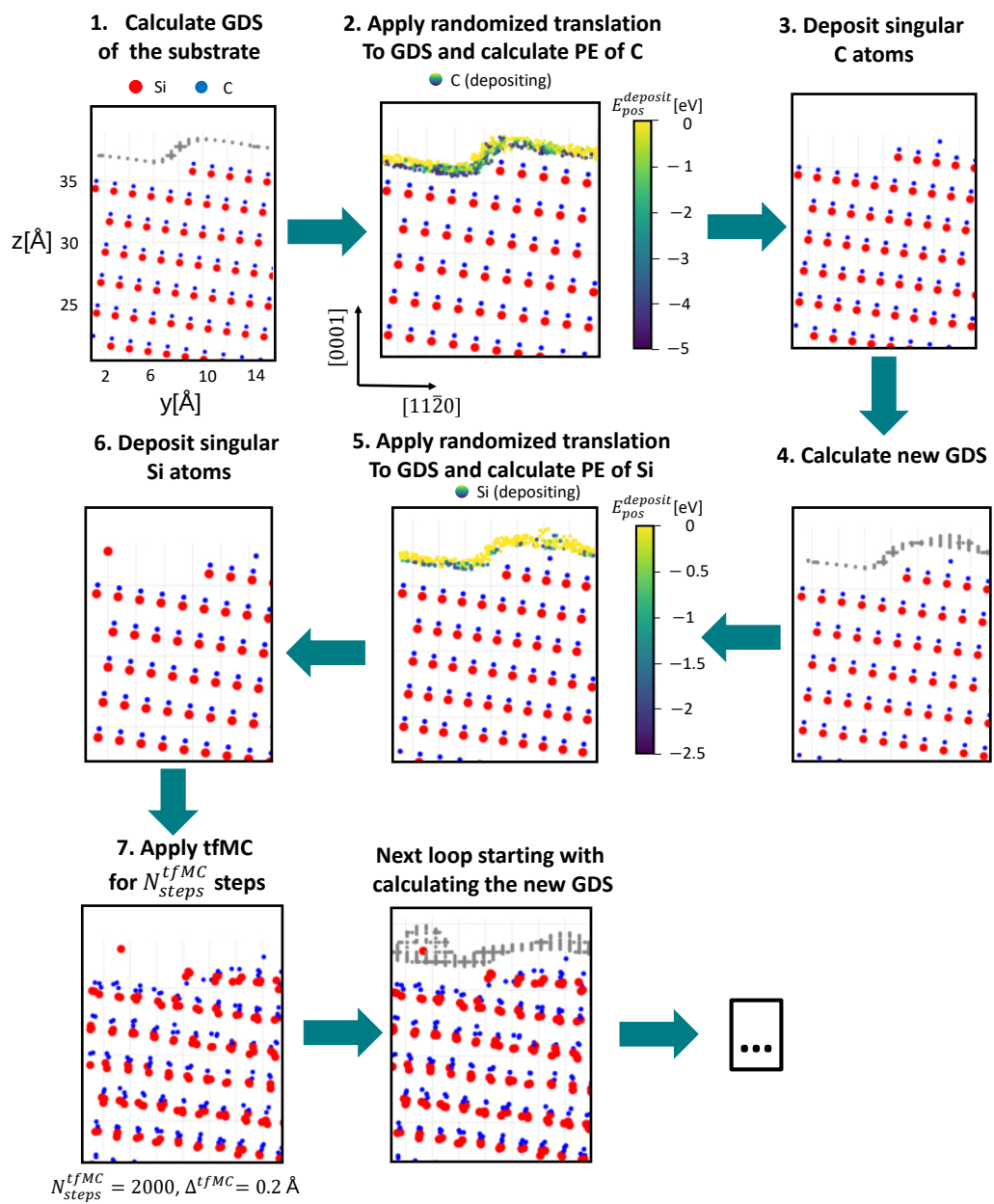

Figure S6: Steps involved in the minimum energy atomic deposition.

## S5. Deposition probability weighting

As can be seen in Fig.S7, sometimes nano pillars form during the growth simulation with the MEAM IP when omitting the probability weighting of the deposition sites. The simulation starts with an initial step-flow growth phase lasting roughly 20 ns, after which this pillar formation is observed. Similar results are gained on flat substrates (Fig.S8). The origins for this behaviour are not fully clarified, but may be specific to this MEAM IP.

When increasing the number of tfMC steps in between deposition steps, the formation of these nano pillars vanishes (Fig.S9). For the $7 \times 7$ stepped substrate using 4000 tfMC steps ($\sim$60 m/h deposition rate) 4H could be stabilized using step flow growth. There is, however, a hexagonal inclusion visible in the middle of the bulk (Fig.S9b), which nucleated on the growth front due to step bunching. A detailed evolution of this stacking fault can be seen in Fig.S10. After 76 ns the four steps of the initial substrate, which are all one atom layer in height have merged together into one. Then the stacking order of 4H is interrupted due to step bunching and a stacking fault is nucleating at the growth front. After 230 ns however the large step has split into multiple steps again and the 4H structure could maintained. Then until 620 ns no further step bunching occurred and the 4H structure was stabilized until the end of the simulation. Increasing the number of tfMC steps further to 8000 ($\sim$30 m/h deposition rate) leads to defect free stabilization of the 4H structure for which no step bunching was observed (Fig.S9c). For a flat $7 \times 7$ substrate, however, even 8000 tfMC steps were insufficient for the prevention of the nano pillar formation (Fig.S9d). This would indicate that, the flatter the substrate is, the more tfMC steps are necessary to prevent these surface artifacts from forming.

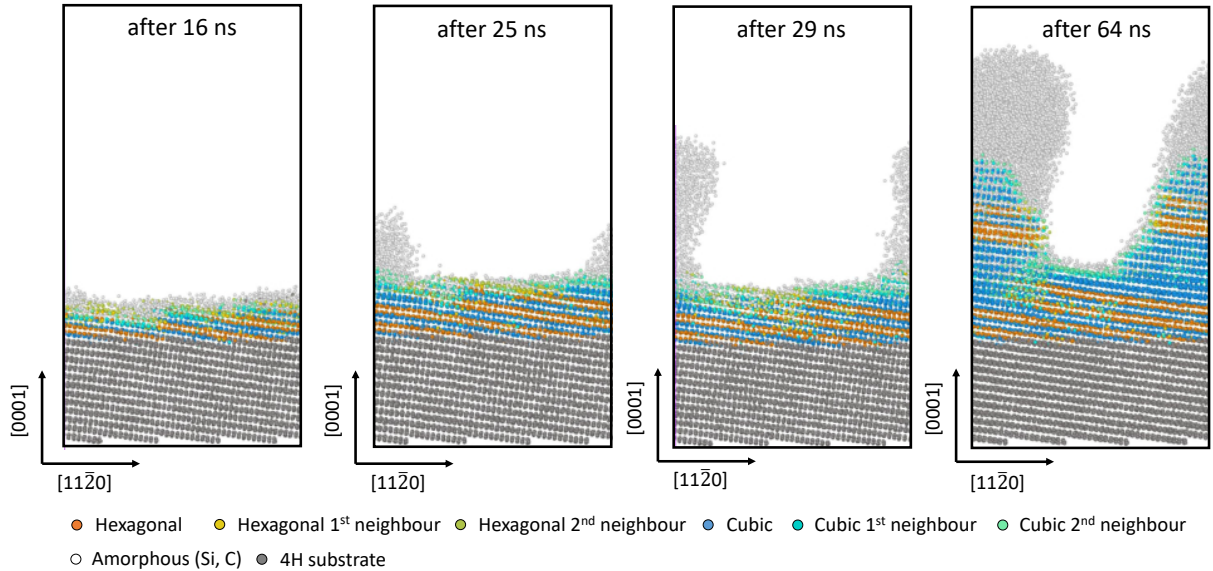

Figure S7: Crystal growth simulation on a stepped 4H C-terminated SiC substrate, without probability weighting in favor of deposition sites lower in z-direction [0001]. For this simulation 1000 tfMC steps have been used in between depositions and the temperature was set to 2500K.

9

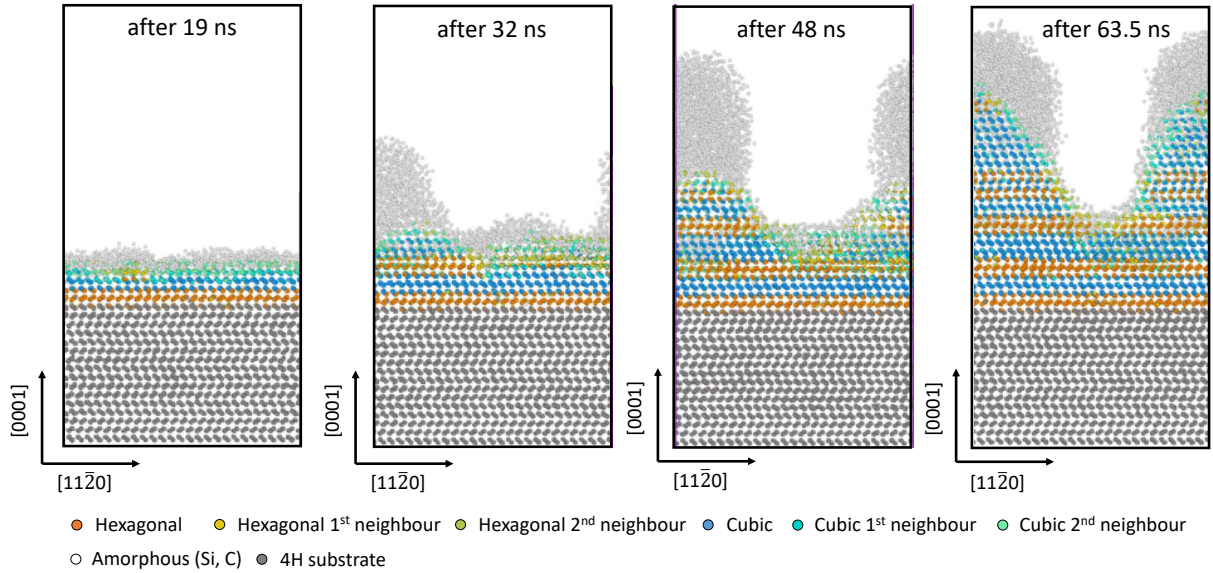

Figure S8: Crystal growth simulation on a flat 4H C-terminated SiC substrate, without probability weighting in favor of deposition sites lower in z-direction [0001]. For this simulation 1000 tfMC steps have been used in between depositions and the temperature was set to 2500K.

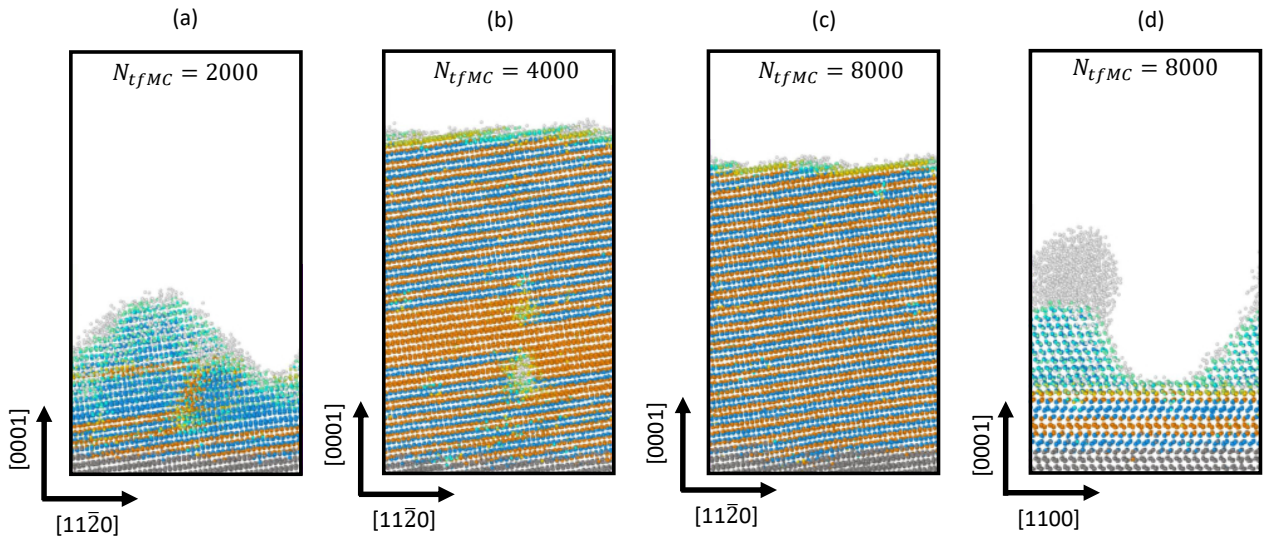

Figure S9: Crystal growth simulation on the 7 × 7 stepped (a-c) and flat (d) substrate using different amounts of tfMC steps $N_{tfMC}$. In these simulations no probability weighting of the deposition sites was applied and the site was randomly chosen after the energy cut-off removed unfavorable deposition sites.

10

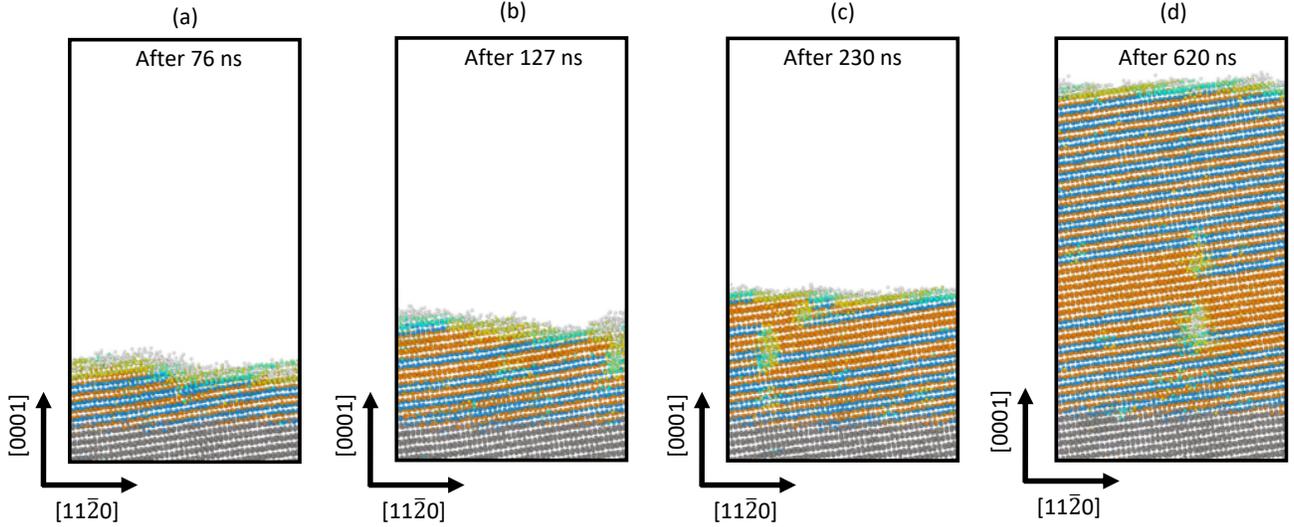

Figure S10: The time evolution of growth simulation using $N_{tfMC} = 4000$.

One way to prevent the formation of nano pillars, as well as, other artifacts, such as deposition of desorbed atoms, is to apply a probability weighting of the potential deposition sites to reduce the probability of depositing atoms far away from the surface. Here the probability weighting was defined in the following way

$$w = (1 - r_d)^{\Delta z - 1} \text{ for } \Delta z > 1 \text{Å}, \tag{1}$$

which decreases the probability for atoms at least 1 Å away from the lowest generated deposition site (in [0001] direction). For our crystal growth simulations we used $r_d = 0.5$ and $r_d = 0.9$. How the value of $r_d$ effects the probability weighting curve is depicted in Fig.S11. On the 7 × 7 stepped substrates, using 1000 tfMC steps, 4H stabilization through step flow grow was achieved (main text Fig.2), for which the growth is similar to the simulation with 8000 tfMC steps, without the probability weighting of the deposition sites (Fig.S9c). We therefore believe that by applying probability weightings of the deposition step a slow down of the growth rates can be achieved. This slow down of the growth rates is especially crucial for the 14 × 14 structure, which has a lower cut-off angle, and therefore more tfMC steps would be necessary for the prevention of unwanted artifacts and the stabilization of 4H, through step flow growth. As can be seen in the main text Fig.3b, with $r_d = 0.5$ and 8000 tfMC, nano pillar formation was prevented, however, the growth rate was too fast to stabilize the 4H structure through step flow growth. Only after increasing the decay rate to $r_d = 0.9$ (main text Fig.3a) straight, evenly spaced steps are observed throughout the growth simulation. When looking at the PEL of Si and C atoms on stepped surfaces (Fig.S12 and Fig.S13), it can be seen that for most IPs the lowest energy positions are on the step, therefore applying a higher probability for deposition in, or near, the step would be naturally achieved by the energy cut-off. For the MEAM IP, however, low energy positions are scattered around the stepped structure and no clear favoring of sites near the step is visible.

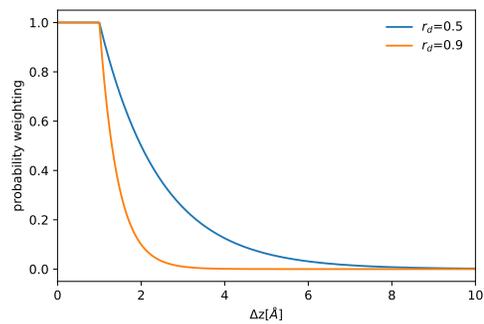

Figure S11: The two probability weightings used for the crystal growth simulations. A value of 1 indicates that no reduction in the probability of choosing the site is applied.

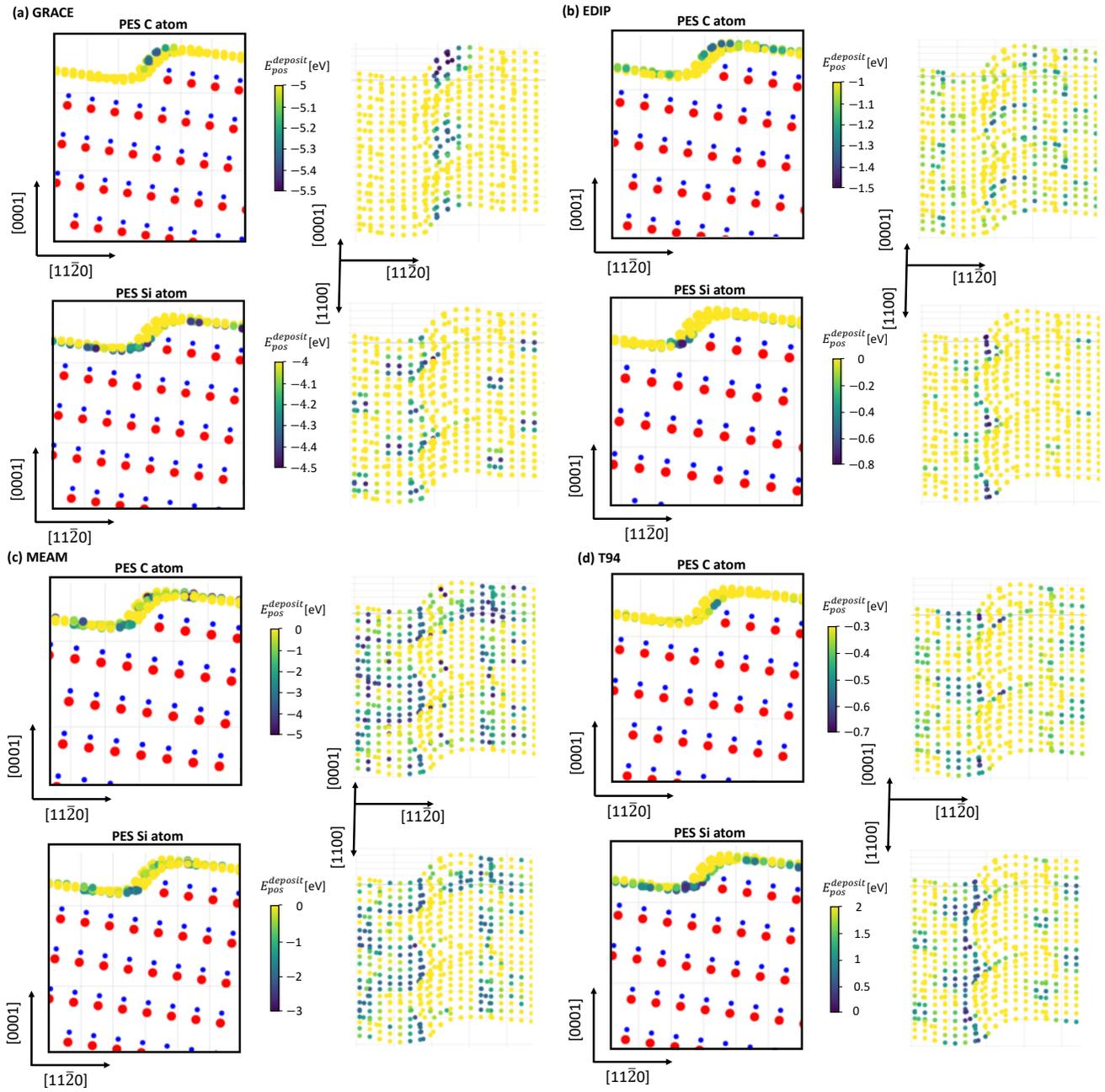

Figure S12: PEL of (a) the GRACE, (b) EDIP, (c) MEAM and (d) T94 IP on stepped surfaces using a C-terminated 4H substrate with 8.1° off-cut angle.

13

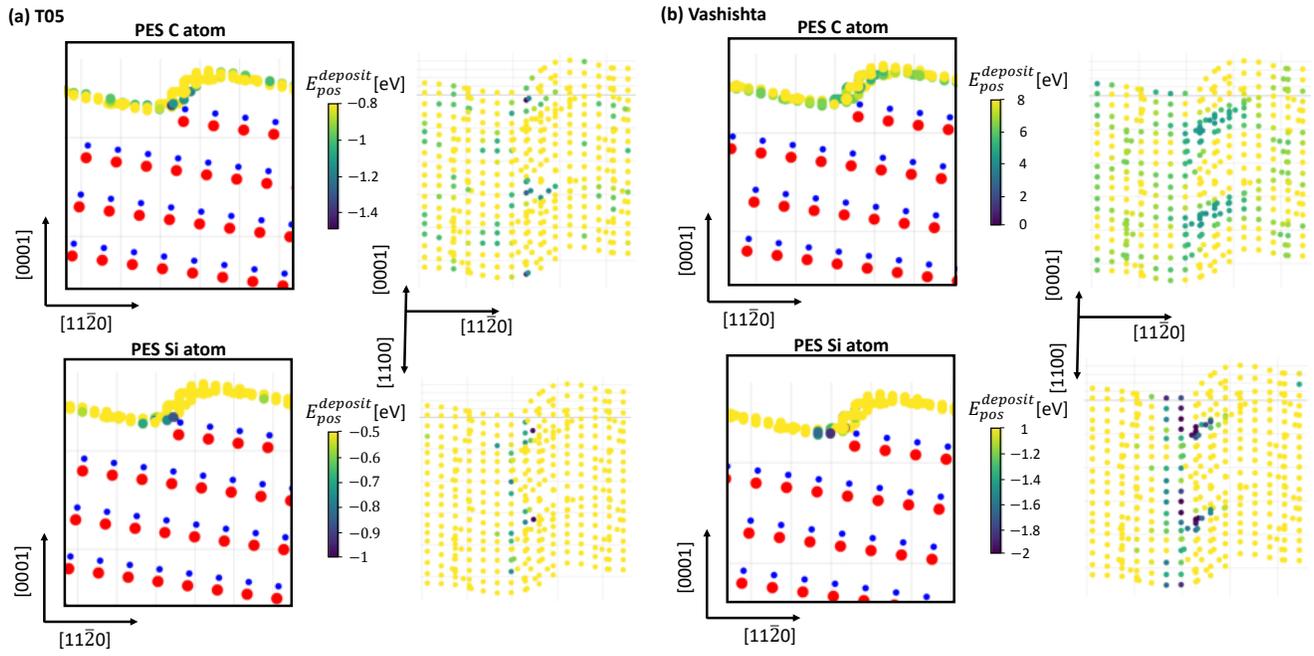

Figure S13: PEL of (a) the T05 and (b) the Vashishta IP on stepped surfaces using a C-terminated 4H substrate with 8.1° off-cut angle.

## S6. Crystal growth simulation on the 14x14 substrate

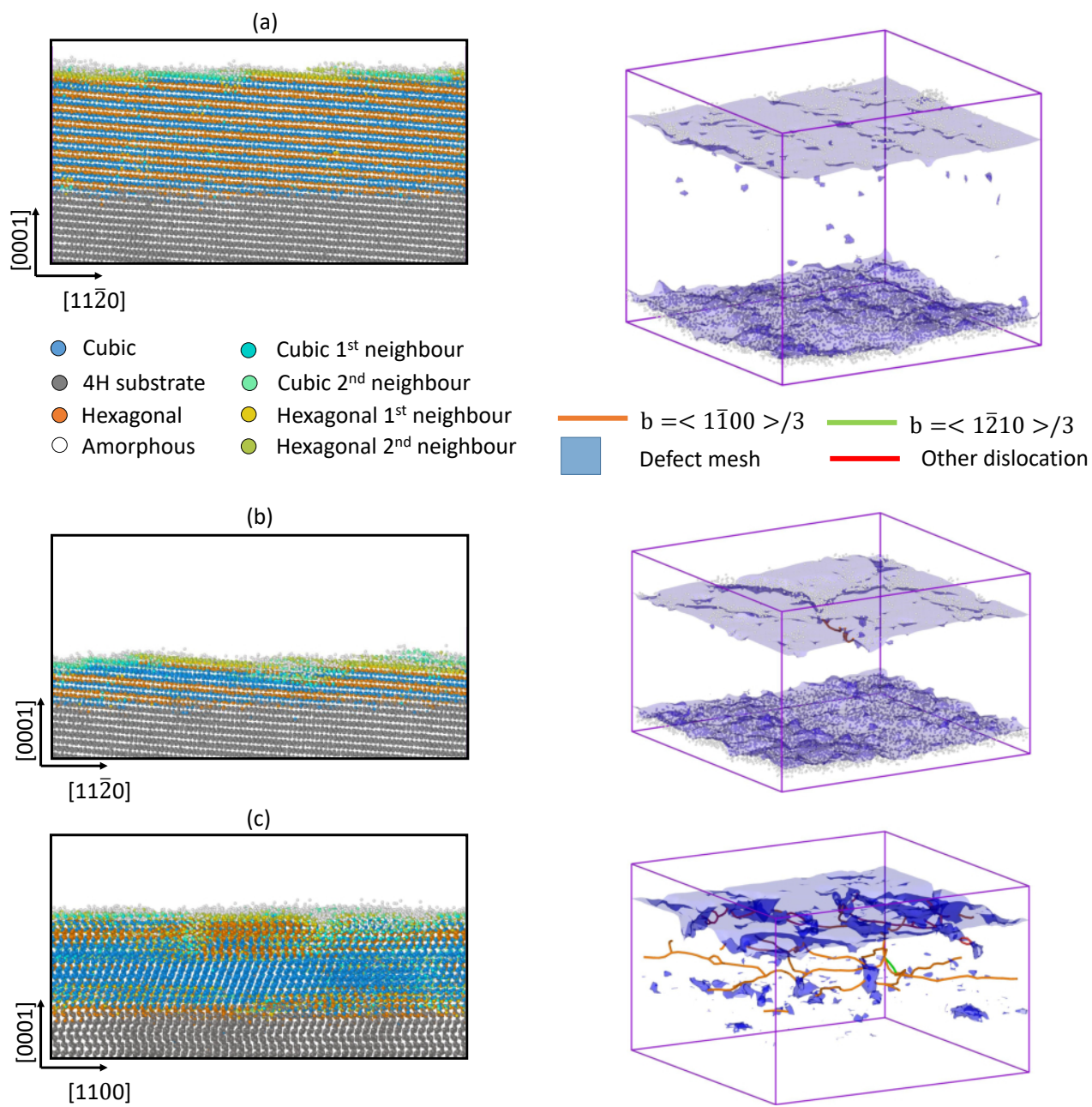

Figure S14: The final crystal structure of the crystal growth simulations on the stepped (a,b) and flat (c) 14x14 4H C-terminated SiC substrates, including their respective dislocation analysis, on the right side.

15

## S7. Crystal growth simulation on the 7x7 substrate using conventional MD

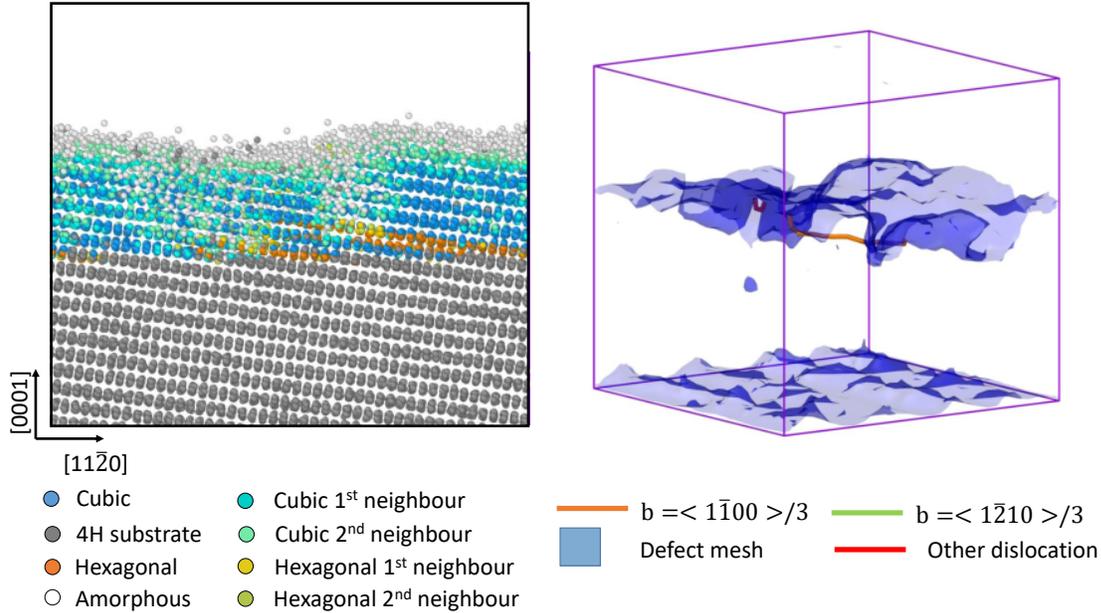

Figure S15: Crystal growth simulation using conventional MD with the MEAM potential on the stepped 7x7 substrate.


\end{document}